# Terraced Compression Method with Automated Threshold Selection for Multidimensional Image Clustering of Heterogeneous Bodies

Jiatong Li[1,2], Gang Li[1,2], Nan Su Su Win[1,2], Ling Lin[1,2,*]

1. Medical School of Tianjin University, Tianjin 300072, China.
2. State Key Laboratory of Precision Measuring Technology and Instruments, Tianjin University, Tianjin 300072, China.

* Corresponding author: Ling Lin (e-mail: linling@tju.edu.cn).

**Abstract:**

Multispectral transmission imaging provides strong benefits for early breast cancer screening. The frame accumulation method addresses the challenge of low grayscale and signal-to-noise ratio resulting from the strong absorption and scattering of light by breast tissue. This method introduces redundancy in data while improving the grayscale and signal-to-noise ratio of the image. Existing terraced compression algorithms effectively eliminate the data redundancy introduced by frame accumulation but necessitate significant time for manual debugging of threshold values. Hence, this paper proposes an improved terrace compression algorithm. The algorithm necessitates solely the input of the desired heterogeneous body size and autonomously calculates the optimal area threshold and gradient threshold by counting the grayscale and combining its distribution. Experimental acquisition involved multi-wavelength images of heterogeneous bodies exhibiting diverse textures, depths, and thicknesses. Subsequently, the method was applied after pre-processing to determine the thresholds for terraced compression at each wavelength, coupled with a window function for multi-dimensional image clustering. The results illustrate the method's efficacy in detecting and identifying various heterogeneous body types, depths, and thicknesses. This approach is expected to accurately identify the locations and types of breast tumors in the future, thus providing a more dependable tool for early breast cancer screening.

## 1. Introduction

Breast cancer poses a serious threat to women's lives and exhibits a high prevalence[1], [2]. From 2010 to 2019, the growth rate was 0.5%[3]. In 2020, the number of new breast cancer cases worldwide surpassed lung cancer for the first time, totaling a staggering 2.26 million people[4]. It is predicted that by 2040, there will be over 3 million new cases and more than 1 million new deaths annually solely due to population growth and aging[5]. Despite being the most prevalent cancer among women worldwide, early detection and treatment can result in a cure rate exceeding 90 percent[6]. Thus, regular screening has become crucial for enhancing breast cancer survival rates in the population.

However, some existing screening technologies, including mammography[7], breast ultrasound[8], [9], breast magnetic resonance imaging[10], [11], and digital breast tomosynthesis[12], are not only costly but also demand specialized personnel for execution, posing challenges to conducting regular screening tests, particularly in resource-limited regions[13], [14], [15], [16]. To mitigate this issue, optical transmission imaging has emerged as a promising screening technique. This technique presents new possibilities for early breast cancer screening due to its safety, convenience, and affordability. Breast tissue is boneless and exhibits good transparency. However, if they harbor tumor tissue, significant neovascularization and hemoglobin are produced around the tumor. As light penetrates the breast tissue, prominent shadows—medically referred to as heterogeneity—emerge at the site of tumor tissue in the acquired transmission image[17]. The multidimensional information carried by the optical image can be utilized to quantify tissue scattering and absorption parameters, facilitating tissue type differentiation and the detection of heterogeneous bodies.

Initially employed for breast disease screening by depicting lumps and vascular shadows[18], [19], optical transmission imaging often yielded inaccurate results[20]. Recently, multispectral imaging—a technique of significant interest—has offered an innovative solution to the challenges of transmission imaging. Due to the varied absorption spectra of different biomolecules at distinct light wavelengths, multispectral light passing through tissue can yield monochromatic spectral images containing absorption spectrum information. Multispectral imaging leverages the absorption and scattering properties of diverse biological tissues at various light wavelengths, offering richer information[21]. Consequently, multispectral transmission imaging, integrating transmission imaging with multispectral imaging, holds significant research value for early breast cancer screening[22], [23], [24], [25], [26]. This advancement not only fosters progress in image processing technology but also furnishes a more precise and dependable method for early breast disease diagnosis.

Despite some studies on multispectral transmission imaging, none have successfully identified the location and type of breast tumors. Moreover, the strong absorption and scattering of light by breast tissue lead to low grayscale and signal-to-noise ratios in multispectral transmission images. To tackle this issue, researchers have proposed a frame accumulation method to enhance image quality[27], [28], [29]. However, due to the redundancy of image data caused by frame accumulation, existing clustering methods fail to effectively detect and distinguish different substances in such images. In response, Li introduced a terrace compression method, utilizing the thought of "digital morphological filtering" to achieve nonlinear compression of image grayscale[30]. Li further refined this method, making it more automated in the implementation process compared to the previous version, requiring only the setting of an area threshold and a gradient threshold[31]. Nevertheless, setting these thresholds still demands significant time for repeated manual threshold debugging and does not guarantee the grayscale range post-compression or the resulting image quality. Moreover, Lin's proposed combination of terrace compression with the window function also achieved only the clustering of different types of heterogeneities[32]. Consequently, this paper proposes an improved terraced compression algorithm. It requires only input the size of the smallest heterogeneous body that is expected to be identified, and can adaptively determine the optimal area and gradient thresholds for the image by counting grayscale and combining its distribution. This study sets sixteen heterogeneous bodies of four types, two depths, and two thicknesses. Transmission images of six wavelengths are collected, followed by frame accumulation and filtering preprocessing for each wavelength image. The method selects the optimal threshold value for each wavelength image to perform gradient compression, supplemented with a window function to preserve required grayscale information. Finally, the GMM clustering algorithm is applied to the multidimensional image data composed of images of different wavelengths for analysis. The results demonstrate that the algorithm effectively distinguishes heterogeneous bodies with varying depths. Moreover, for heterogeneous bodies with differing thicknesses, the algorithm categorizes them based on type and groups those of the same type with varying thicknesses together. Additionally, it can also differentiate the thickness of heterogeneous bodies to a certain degree, which proves the validity of the threshold selection, and it is expected to play an important role in clinical experiments to achieve the precise identification of breast tumor location and type.

## 2. Theory

### 2.1 Clustering of Multi-wavelength Images

The clustering principle of multi-wavelength transmission images is based on the fact that different tissues show different absorption effects at distinct wavelengths, and these differences serve as the characteristic information of each sample during clustering. Conventional image clustering uses the grayscale values of grayscale images as sample data, which belongs to one-dimensional clustering. However, in clustering multi-wavelength images, each pixel is treated as a sample, and the grayscale values at each wavelength are considered, resulting in multidimensional clustering.

A multi-wavelength image data of size M×N is illustrated in Table. 1, where $g_{\lambda 1}$, $g_{\lambda 2}$, $g_{\lambda 3}$…$g_{\lambda n}$ represents the pixel value of the pixel point at the corresponding wavelength. If the grayscale value of the cluster center at each wavelength is $g_1, g_2, g_3$…$g_n$, the formula for calculating the distance during clustering is shown in equation (2-1).

Table. 1 Schematic Diagram of Multi-wavelength Image Data Initialization

| Pixel point of wavelength | $\lambda 1$ | $\lambda 2$ | $\lambda 3$ | … | $\lambda n$ |
|---|---|---|---|---|---|
| 1 | $g_{\lambda 1}$ | $g_{\lambda 2}$ | $g_{\lambda 3}$ | … | $g_{\lambda n}$ |
| 2 | $g_{\lambda 1}$ | $g_{\lambda 2}$ | $g_{\lambda 3}$ | … | $g_{\lambda n}$ |
| … | … | … | … | … | … |
| M×N | $g_{\lambda 1}$ | $g_{\lambda 2}$ | $g_{\lambda 3}$ | … | $g_{\lambda n}$ |

$$D = \sqrt{(g_{\lambda 1} - g_1)^2 + (g_{\lambda 2} - g_2)^2 + (g_{\lambda 3} - g_3)^2 + \cdots + (g_{\lambda n} - g_n)^2} \quad (2-1)$$

## 2.2 Terrace Compression Method Based on Automatic Threshold Selection of Grayscale Distribution

This paper proposes an improved terraced compression method to achieve adaptive threshold selection. Inputting the minimum expected area of heterogeneous identification and analyzing the grayscale distribution in the image enables the determination of the optimal gradient and area thresholds for this image based on this analysis. This innovation addresses the threshold setting challenge, significantly reduces debugging costs, and improves the accuracy and efficiency of recognizing heterogeneous bodies. Below is an example detailing the specific steps of the method for obtaining the threshold value for the preprocessed image at a wavelength of 410 nm.

Step 1: Generate the original 3D grayscale distribution map of the image for observation, as depicted in Fig. 1. The x-axis and y-axis represent the coordinate axes of the pixel points in the original image, while the z-axis indicates the grayscale value of the pixel points at the respective coordinates. The grayscale values in the image range from 611 to 28855.

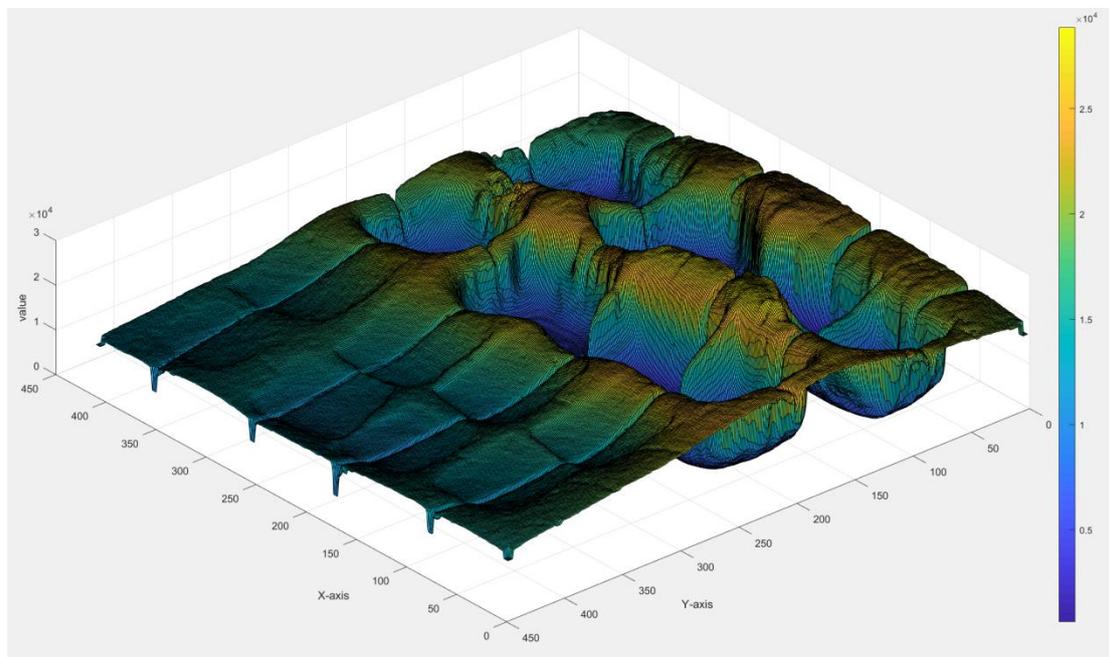

Fig. 1 Original 3D Grayscale Distribution of the Preprocessed Image at 410 nm

Step 2: The pixels are uniformly divided into 35 segments based on their grayscale distribution range. The three-dimensional grayscale distribution is illustrated in Fig. 2, and the grayscale histogram is depicted in Fig. 3. The distribution results are as follows:

Segment 1 covers a grayscale value distribution range from 611.0 to 1418.0, with 5286 pixels falling within this range.

Segment 2 covers a grayscale value distribution range from 1418.0 to 2225.0, with 4674 pixels falling within this range.

Segment 3 covers a grayscale value distribution range from 2225.0 to 3032.0, with 6022 pixels falling within this range.

Similarly, Segment 35 covers a grayscale value distribution range from 28048.0 to 28855.0, with 613 pixels falling within this range.

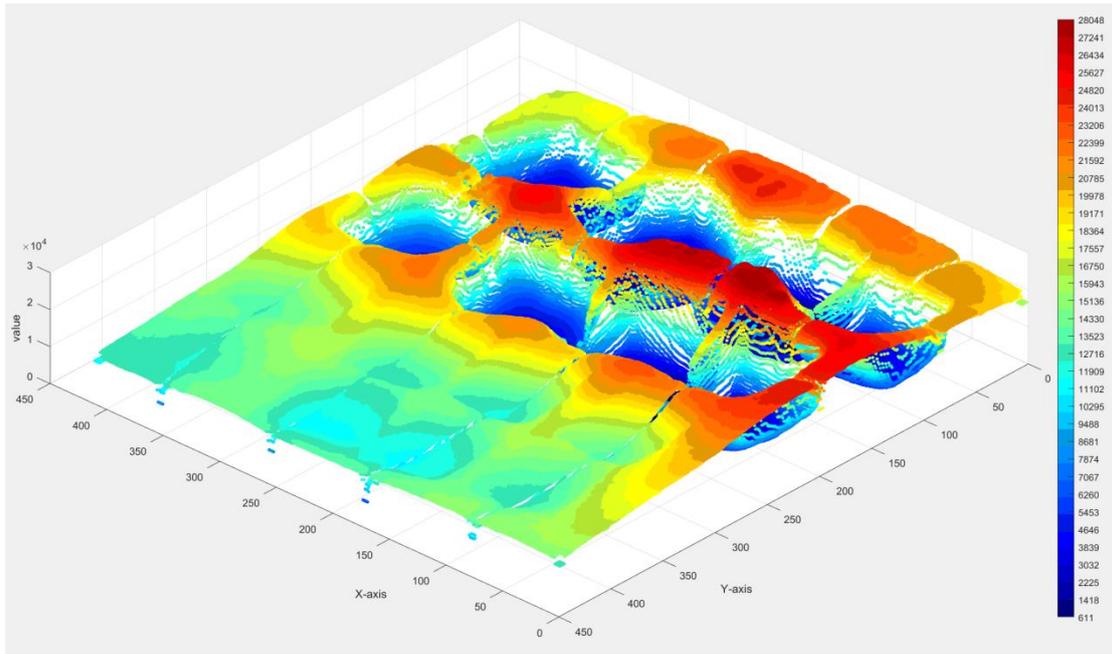

Fig. 2 3D Grayscale Distribution after Segmentation

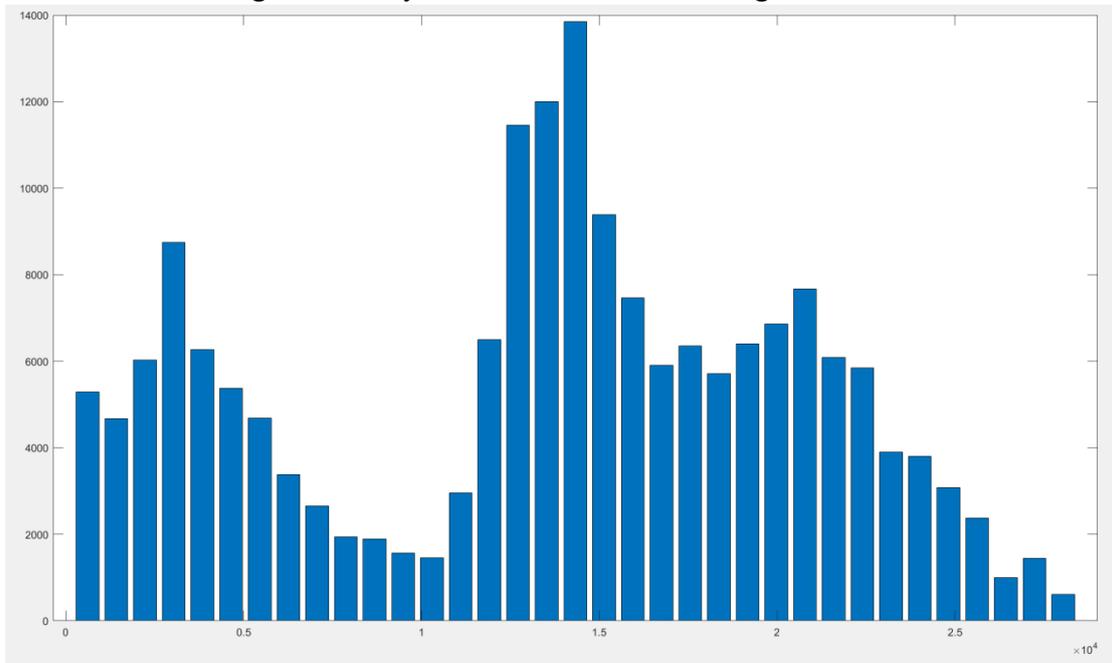

Fig. 3 Histogram of Grayscale after Segmentation

Step 3: If a segment contains fewer pixels than one percent of the maximum grey value (28855), it is merged with neighboring segments having a smaller pixel count until each segment contains more than one percent of the maximum grey value. The result of this step in this example is the same as the second step.

Step 4: Utilizing the grey level histogram obtained in Step 3, a demarcation segment is identified to distinguish between the background and the heterogeneous body. When the heterogeneous body occupies a smaller area of the image, the segment with the largest number of pixel points is selected. Since the dataset of this paper comprises 16 heterogeneous bodies, occupying a substantial portion of the image, the demarcation segment is chosen as the one with the largest number of pixel points among the last three-fifths of segments. In this context, it corresponds to segment 18, featuring a grey value distribution range of [14330.0, 15136.0).

Because the absorption of the heterogeneous body exceeds that of normal tissue, resulting in lower grayscale values in the image, all the grayscale value distribution ranges above the demarcation segment are regarded as the grayscale value ranges of the heterogeneous body distribution, ranging from [611.0, 14330.0), while those below it are regarded as the grayscale value ranges of the background distribution, ranging from [15136.0, 28855.0).

Step 5: Based on the grey value range of the heterogeneous body distribution obtained in Step 4, identify the segment with the smallest number of pixels. Consider the grey value range between the next segment of this segment and the upper two segments above the demarcation segment as the more accurate range of the heterogeneous body distribution. Consider this range's size as the gradient threshold.

Owing to the varied depths of heterogeneous bodies in our dataset, their grayscale performance exhibits great differences. Specifically, the grayscale range of the heterogeneous body distribution is broader in the shallower depth (the first two rows), potentially resulting in suboptimal compression for the heterogeneous body in the deeper depth (the last two rows) if the gradient threshold is based solely on the former. Therefore, the selection of the gradient threshold should be based on the range of the grayscale distribution of the heterogeneous body in the deeper depth. Upon examining the original 3D grayscale distribution maps of the six wavelengths, we observe that the grayscale values' range for the heterogeneity distribution in the deeper depth all exceeds 5000. Consequently, an additional criterion must be introduced: within the range of grayscale values for the heterogeneity distribution, identify the segment with a grayscale value exceeding 5000 and containing the fewest pixels. Here, segment 13 denotes the more accurate range for the heterogeneous body distribution as [11102.0, 13523.0), with a gradient threshold of 2421, as depicted in Fig. 4.

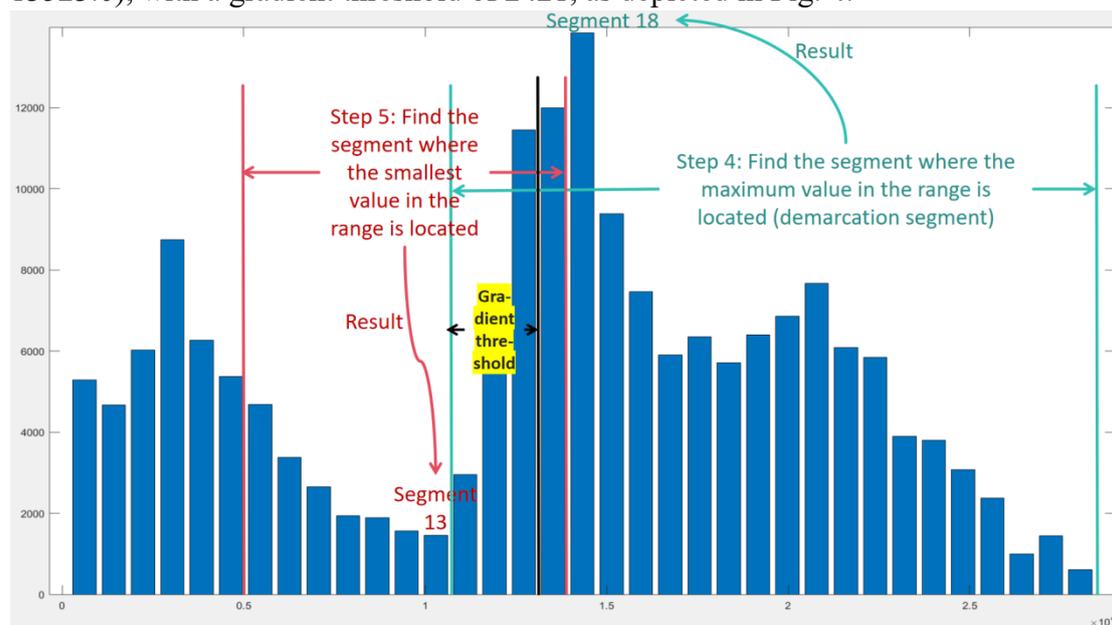

Fig. 4 Schematic of Gradient Threshold Acquisition

Step 6: The area threshold should be equal to or greater than the desired size of the recognized heterogeneous body. Thus, the initial area threshold is set to the input area size, and the initial gradient threshold is set to the value obtained in Step 5.

As the area threshold increases, the grayscale range of the image decreases post-compression. While the gradient threshold decreases, the grayscale range of the image either remains the same or decreases post-compression. Hence, if the post-compression grayscale range of the image exceeds 255, the gradient threshold remains constant, and the area threshold is increased to 1.2 times its previous value for re-compressing the image. This iterative process continues until the maximum grayscale value of the compressed image drops below 255, signifying the finalization of both the area threshold and gradient threshold.

**2.3 Grey Scale Window Transformation**

Grayscale window transform is a window function-based grayscale transformation method that utilizes segmented function implementation. By selecting windows of various sizes to operate on the grayscale, its objective is to preserve the grayscale region of interest while eliminating unnecessary grayscale details. The selection of different window sizes significantly impacts image processing. Linearly stretching the grey scale region of interest can enhance detailed information within it. Fig. 5 illustrates the window function. Specifically, the function maps grey levels larger than the window to the maximum value, those smaller than the window to 0, and linearly transforms grey levels within the window to achieve stretching or compression within the region.

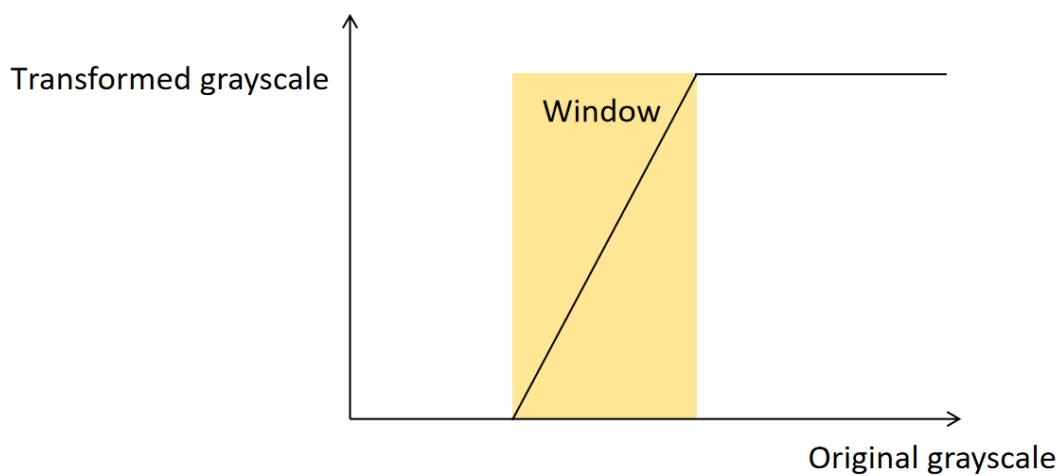

Fig. 5 Window Function

## 3. Experiments

This study designed a comprehensive set of breast mimicry experiments to validate the proposed method's effectiveness. The experiments involved acquiring transmission images at six wavelengths and applying various processes, including preprocessing, terraced compression, window function transformation, and multi-wavelength image clustering. The experimental process primarily comprises three aspects: experimental setup, image acquisition, and image processing. Fig. 6 illustrates the specific experimental flow.

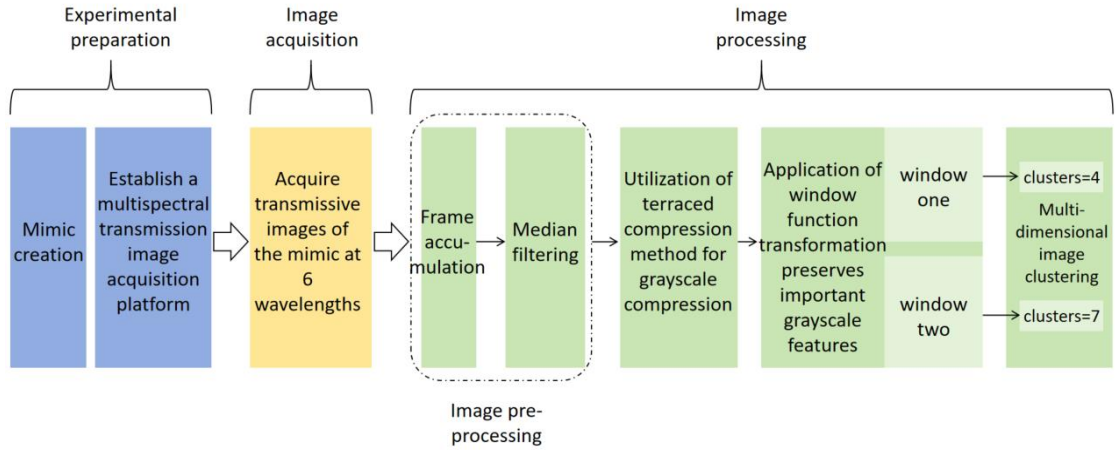

Fig. 6 Experimental Flow Chart

## 3.1 Experimental Setup

The multispectral transmission image acquisition system, as depicted in Fig. 7, includes nine identical six-in-one LED beads emitting wavelengths ranging from 410 nm, 450 nm, 500 nm, 590 nm, 660 nm, 734 nm, equipped with a built-in fan for heat dissipation. Utilizing the Halian Nuctech industrial camera (Model: MV-CA016-10UC), the image acquisition segment captures data, while the subsequent image processing phase is carried out using a computer equipped with MATLAB R2019a and Python 3.8. Given that it is currently in the methodological research stage, a mimic body was used to simulate the breast experiment in order to obtain transmission images similar to those of breast tissue. This mimic body comprised a fat milk mixture injected with water at a ratio of 1:1150 to simulate normal breast tissue. Positioned at the center of the mimics were fat meat, potato, carrot, and lean meat pieces to simulate heterogeneous structures in the mammary gland. The dimensions of these structures varied, and their positional distributions are shown in the figure below, with the first and third rows measuring 10 mm × 10 mm × 10 mm, and the second and fourth rows measuring 10 mm × 10 mm × 5 mm. Additionally, the first two rows of Potato, Carrot, Lean Meat, and Fatty Meat were situated on the same plane and placed near the camera, while the last two rows were located on the same plane and positioned closer to the light source.

The control mimic comprised clear water, with identical heterogeneous bodies placed in the same locations as those in the mimic to acquire template images necessary for evaluating segmentation effects.

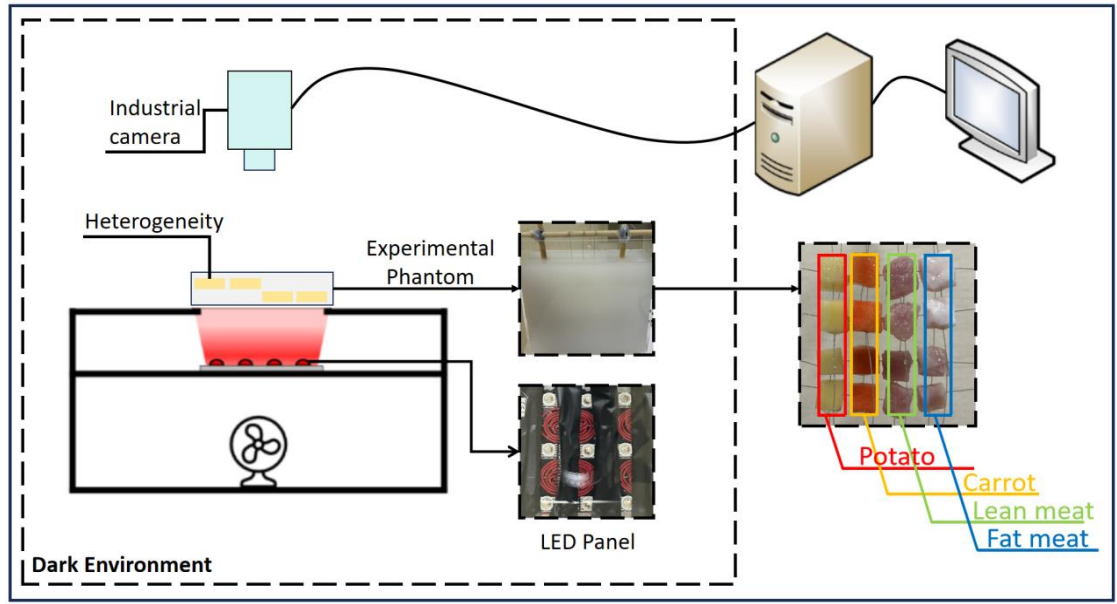

Fig. 7 Experimental Setup Diagram

**3.2 Image Acquisition**

(1) Configure the image resolution to 480×444 and set the acquisition frame rate to 30 fps. Adjust the distance between the camera, the control mimic, and the light source to ensure that the light shines evenly on the mimic to prevent overexposure. Adjust the focus of the camera for clear and comprehensive image capture. Utilize a blackout cloth to shield the entire experimental setup from ambient light interference.

(2) Power on the light source and capture the template image. Then, inject the fat milk solution into the control mimic at the specified ratio and continue image collection after complete mixing. Collect 700 frames of images at each wavelength. Upon completing acquisition at one wavelength, replace only the light source while maintaining the camera and mimic positions unchanged. Proceed with image acquisition. The first frame of acquired images at each wavelength is depicted in Fig. 8.

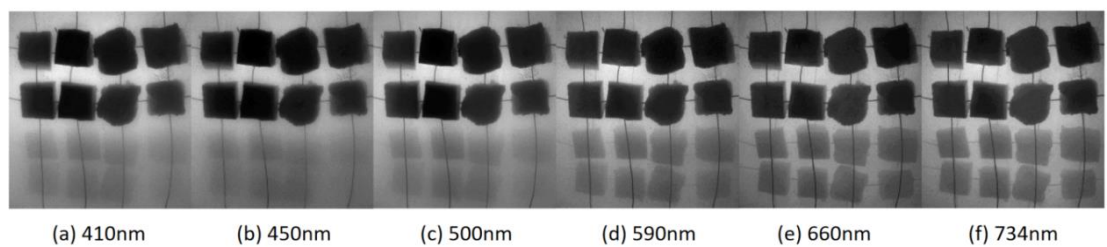

Fig. 8 The First Frame of the Captured Original Image

**3.3 Image Processing**
**3.3.1 Image Pre-processing**

(1) Frame Accumulation: Transmission images collected at each wavelength undergo frame accumulation. Specifically, 700 images are accumulated per wavelength, and the outcomes are illustrated in Fig. 9 (a1)-(f1).

(2) Filtering: The presence of minor noise, often attributed to bubbles, necessitates the application of median filtering to enhance image smoothness. Median filtering, a non-linear filter, operates by sorting the grey values of pixel points and their neighboring points and replacing the point's grey value with the median. In this study, a 5×5 window is chosen for median filtering, and the filtered result is displayed in Fig. 9 (a2)-(f2).

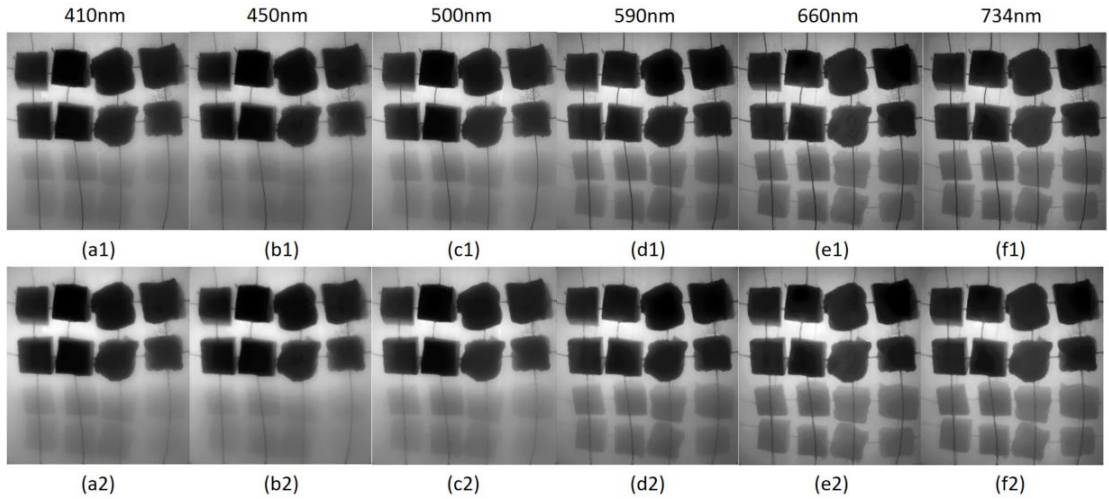

Fig. 9 Preprocessing Result –
Top row (a1)-(f1) shows images after frame accumulation
Bottom row (a2)-(f2) displays filtered images

**3.3.2 Terraced Compression**

After preprocessing, wherein noise is effectively reduced, the terraced compression process is initiated. Utilizing the threshold selection method outlined in Section 2.2, the most suitable gradient threshold and area threshold are determined by inputting the minimum size of the anticipated recognized heterogeneous bodies. The specific thresholds used for compression are detailed in Table. 2. The compressed images after terraced compression for each wavelength are illustrated in Fig. 10 (a1)-(f1).

Table. 2 Threshold Acquisition for Terraced Compression Method

| Wavelengths | Gray-scale range before compression | Input parameters | Obtained gradient threshold | Obtained area threshold | Gray-scale range after compression |
|---|---|---|---|---|---|
| 410nm | 611-28855 | 900 | 2421 | 900 | 1-126 |
| 450nm | 1-21121 | 900 | 1810 | 900 | 1-119 |
| 500nm | 0-30686 | 900 | 2630 | 900 | 1-128 |
| 590nm | 1506-22045 | 1225 | 1761 | 1225 | 1-99 |
| 660nm | 1401-17365 | 900 | 1368 | 900 | 1-122 |
| 734nm | 1489-19910 | 900 | 2105 | 900 | 1-134 |

**3.3.3 Window Function Transformation**

Heterogeneous bodies generally exhibit higher absorbance levels compared to normal tissues, resulting in lower grayscale values for heterogeneous regions and higher grayscale values for the background in transmission images. The image processing algorithm selects two window sizes based on the grayscale values of the compressed images at various wavelengths and the grayscale values of the heterogeneous bodies. Window function one preserves low grayscale information corresponding to heterogeneous bodies while adjusting high grayscale values of the background to the same grey scale. Window function two retains information related to lighter-depth heterogeneous bodies and adjusts the grayscale of deeper-depth heterogeneous bodies and background to the same grey scale. The specific window sizes are detailed in Table. 3. The resulting images are depicted in Fig. 10 (a2-f2) and (a3-f3).

Table. 3 Selection of Window Function Range

| Wavelengths | 410nm | 450nm | 500nm | 590nm | 660nm | 734nm |
|---|---|---|---|---|---|---|
| Grayscale range after compression | 1-126 | 1-119 | 1-128 | 1-99 | 1-122 | 1-134 |
| Grayscale range of window one | 0-95 | 0-80 | 0-89 | 0-87 | 0-95 | 0-95 |
| Grayscale range of window two | 0-55 | 0-45 | 0-40 | 0-38 | 0-55 | 0-50 |

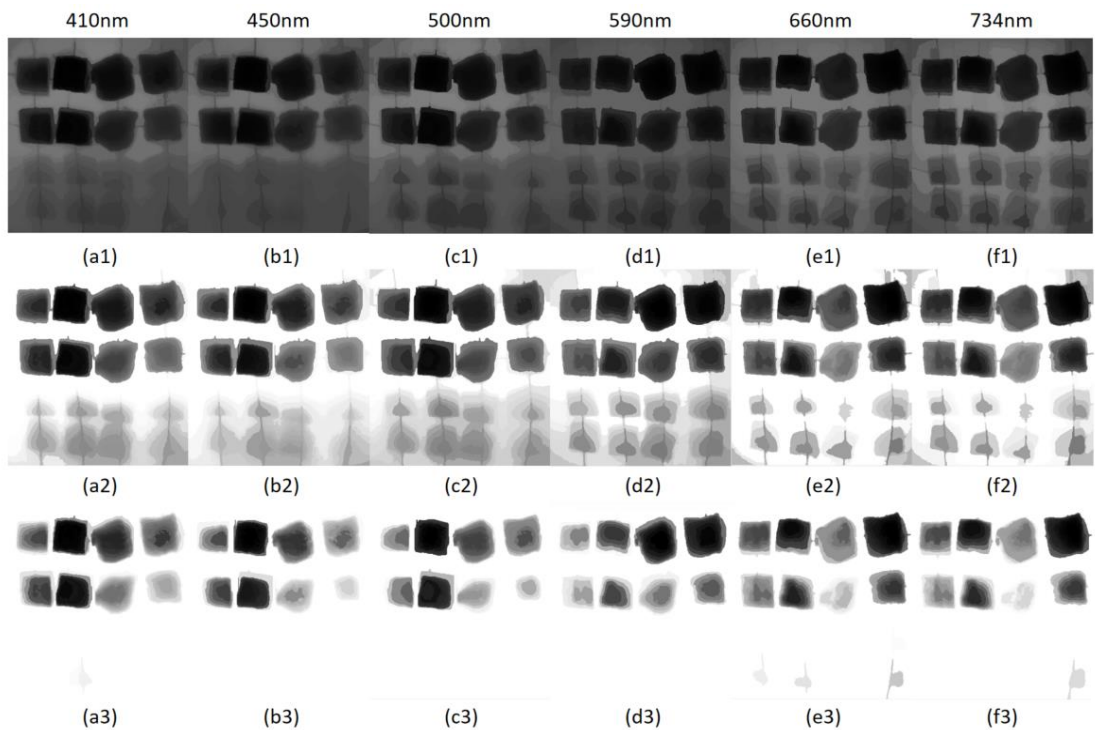

Fig. 10 Terraced Compressed Images and Window Function Images –
Top row (a1)-(f1): Terraced compressed images of each wavelength
Middle row (a2)-(f2): Images of window function one for each wavelength
Bottom row (a3)-(f3): Images of window function two for each wavelength

### 3.3.4 Multi-dimensional Image Clustering

Each wavelength's image is represented as a 2D matrix of dimensions 480×444. Before clustering, the image data requires organization. A reshape operation is executed on the image matrix for each wavelength, converting it into a 1 × 213120 column vector with six columns representing the six wavelengths. Subsequently, these six columns are concatenated to form a 6 × 213120 matrix, serving as the input data for multi-dimensional image clustering. This process yields 213120 groups of clustered data, each characterized by 6-dimensional feature measures. The final classification result is derived by determining the number of clusters. In this study, the Gaussian Mixture Model (GMM) method is utilized for multi-dimensional image clustering.

## 4. Results and Discussion

To validate the efficacy of multi-dimensional image clustering, integrating the terraced compression method based on automatic threshold selection of grayscale distribution and window function for the differentiation of heterogeneous bodies, this section conducts a comparative analysis with conventional clustering techniques, including K-means, K-means++, Mean-shift, and GMM. The overall clustering effectiveness and the accuracy of heterogeneous body segmentation using the two window functions are evaluated.

As the outcome of heterogeneous body clustering segmentation is sensitive to the number of clusters, it's crucial to establish an equal number of clusters for each method during comparison. The clustering outcomes are represented using distinct colors to denote various categories of pixel points and are illustrated in Fig.

Fig. 11 illustrates the template image post-binarization, where each heterogeneous body is identified by an ordinal number. Following the binarization of the template image and images acquired by each clustering method, the images containing only a single heterogeneous body at the same size and location are individually isolated. Subsequently, segmentation accuracy is evaluated by calculating the Dice Coefficient (Dice) for each heterogeneous body between the template and the results obtained from each clustering method.

The Dice coefficient is a crucial evaluation metric employed to assess the similarity of two sets and is widely applied in medical image segmentation evaluation. It is calculated as twice the intersection of two sets divided by the sum of the two sets, as demonstrated in equation (4-1). The Dice coefficient ranges from 0 to 1, with higher values indicating greater similarity between the sets, thus indicating more accurate segmentation. A Dice coefficient of 0 implies no similarity, while a value of 1 indicates identical sets.

$$Dice = \frac{2|A \cap B|}{|A| + |B|} = \frac{2TP}{2TP + FP + FN} \qquad (4-1)$$

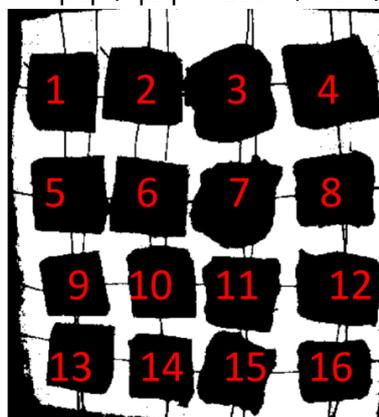

Fig. 11 Template Image after Binarization

## 4.1 Terrace Compression + Window Function One + GMM

The number of clusters for each clustering method is set to 4. The clustering results are depicted in Fig. 12, and the Dice coefficients for each heterogeneous body under different methods are calculated and presented in Table. 4. These results demonstrate that K-means, K-means++, Mean-shift, GMM, and the method presented in this paper effectively distinguish heterogeneous bodies with varying depths. Notably, the method presented in this paper exhibits the most favorable segmentation outcomes, achieving an average Dice coefficient of 0.5982.

The GMM method is the least effective in segmenting the heterogeneous bodies at deeper depths (the last two rows), as it fails to recognize the contours and positions of these bodies. Similarly, both the K-means and K-means++ methods yield comparable outcomes, with only partial recognition of the heterogeneous bodies in the third row, and the heterogeneous bodies in the fourth row being grouped in the same category with the adjacent boundaries. The mean-shift method fares even worse; the heterogeneous body in the fourth row merged with the heterogeneous body in the third row, rendering its contour and position unrecognizable. It is evident that the aforementioned four methods are ineffective for segmenting heterogeneous bodies at deeper depths.

Compared to the previous four results, the method presented in this paper exhibits a significant enhancement in delineating the edges of the heterogeneous bodies at deeper depths (the last two rows). With the exception of the 13th heterogeneous body, optimal segmentation outcomes were achieved for each remaining heterogeneous body, marked by the largest Dice values and an average improvement of at least 24.35%. Additionally, employing the GMM method for clustering, gradient compression using the thresholds selected by the method proposed in this paper not only improves the Dice coefficient by 29.26% overall but also enhances the recognition rate of deeper heterogeneous bodies by 167.6% compared to no gradient compression. This underscores the efficacy of the proposed method in identifying heterogeneous bodies, particularly those with low grayscale values situated at deeper positions, thereby affirming the effectiveness of terrace compression and the accuracy of threshold selection.

Unlike the outcomes of the preceding four clustering methods, the clustering outcomes of the method presented in this paper also encompass information regarding the thickness of the heterogeneous body. The decreased thickness of the heterogeneous body in the fourth row, compared to the third row, leads to reduced light transmission through the wires within the heterogeneous body, as depicted in the graph. Future advancements, coupled with pattern recognition techniques, are expected to facilitate the identification of heterogeneous bodies with diverse thicknesses from a multidimensional standpoint.

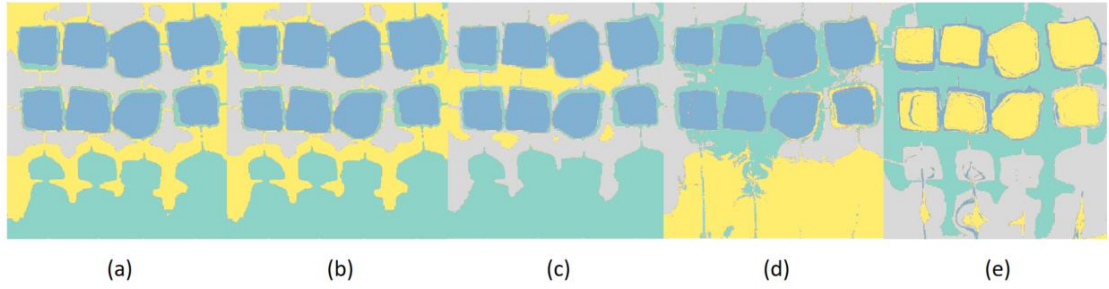

Fig. 12 Clustering Results - (a) K-means; (b) K-means++; (c) Mean-shift; (d) GMM; (e) Terrace Compression Method + Window Function I + Clustering

Table. 4 Dice Values of Heterogeneous Bodies under Different Methods

|  | K-means | K-means++ | Mean-shift | GMM | The method presented in this paper/ Terrace Compression Method + Window Function I + GMM |
|---|---|---|---|---|---|
| 1 | 0.5866 | **0.5869** | 0.5786 | 0.5657 | 0.5058 |
| 2 | **0.8614** | **0.8614** | 0.8564 | 0.7376 | 0.5758 |
| 3 | 0.9077 | **0.9083** | 0.9035 | 0.9029 | 0.816 |
| 4 | 0.8849 | **0.8851** | 0.8802 | 0.8493 | 0.7968 |
| 5 | 0.625 | **0.6261** | 0.6173 | 0.6079 | 0.5358 |
| 6 | 0.801 | **0.8012** | 0.7977 | 0.7169 | 0.6466 |
| 7 | 0.8286 | 0.829 | 0.8234 | **0.8425** | 0.8221 |
| 8 | 0.6616 | 0.6623 | 0.6504 | 0.6221 | **0.6982** |
| 9 | 0.5427 | 0.5439 | 0.5723 | 0.5478 | **0.7652** |
| 10 | 0.4553 | 0.4554 | 0.4661 | 0.2132 | **0.5082** |
| 11 | 0.463 | 0.4613 | 0.4009 | 0.2284 | **0.5937** |
| 12 | 0.4908 | 0.4926 | 0.5059 | 0.391 | **0.6037** |
| 13 | **0.4973** | 0.4954 | 0.3782 | 0.0564 | 0.3629 |
| 14 | 0.3053 | 0.2914 | 0.0885 | 0.1227 | **0.3689** |
| 15 | 0.3689 | 0.3575 | 0.0599 | 0.0008 | **0.6614** |
| 16 | 0.2345 | 0.2282 | 0.0783 | 0 | **0.3108** |
| Overall Average | 0.5947 | 0.5929 | 0.5411 | 0.4628 | **0.5982** |
| Mean Values for Heterozygotes 9-16 | 0.4197 | 0.4157 | 0.3188 | 0.195 | **0.5219** |

Note: Bolded numbers represent the maximum Dice value.

**4.2 Terrace Compression + Window Function Two + GMM**

Each clustering method is configured with 7 clusters. The clustering results are illustrated in Fig. 13, and the Dice coefficients for each heterogeneous body obtained under different methods are presented in Table. 5.

As discussed in Section 4.1, which focused on heterogeneous body segmentation in deeper depths (the last two rows), this section investigates the results obtained at shallower depths (the first two rows).

While the mean-shift and GMM methods yielded superior segmentation results with the highest average Dice coefficient, they both misclassified heterogeneous bodies. Specifically, the mean-shift method erroneously grouped carrots, some potatoes, thicker lean meats, and fattier meats into the same category. Likewise, the GMM algorithm inaccurately categorized carrots and lean meat together, as well as potatoes and fatty meat together, while also failing to recognize potatoes and fatty meat in the fourth row.

By combining the improved terrace compression method proposed in this paper with window function II, the clustering results are displayed in Fig. 13 (e). In contrast to the K-means and K-means++ methods, the proposed approach successfully distinguishes heterogeneous bodies based on categories without significantly reducing the Dice coefficients. Additionally, it categorizes heterogeneities of the same species with varying thicknesses into the same category. Furthermore, a notable reduction in the edge effect is observed for the second row of heterogeneous bodies, which have smaller thicknesses compared to those in the first row. Similarly, future efforts, coupled with pattern recognition techniques, are expected to enhance the recognition of heterogeneous bodies with diverse thicknesses from a multidimensional perspective.

In summary, compared with the four traditional clustering methods of K-means, K-means++, Mean-shift, and GMM, the terraced field compression method proposed in this paper, based on automatic threshold selection of grayscale distribution, effectively distinguishes heterogeneous bodies when combined with the window function for clustering. This results in clearer boundaries of the heterogeneous bodies, thereby enhancing their recognition.

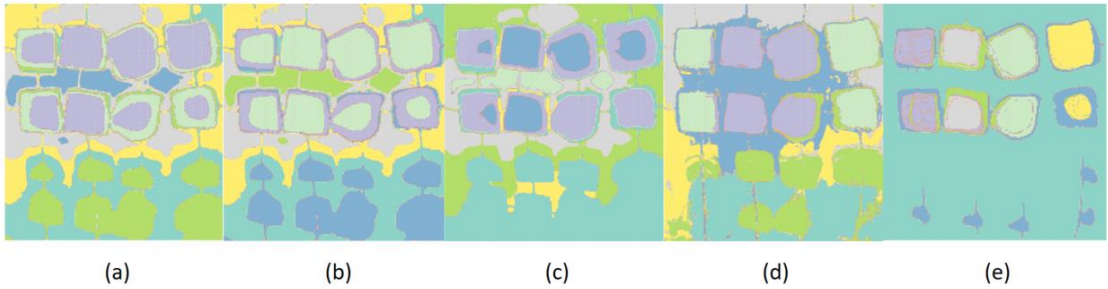

Fig. 13 Clustering Results - (a) K-means; (b) K-means++; (c) Mean-shift; (d) GMM; (e) Terrace Compression Method + Window Function II + Clustering

Table. 5 Dice Values of Heterogeneous Bodies under Different Methods

|   | K-means | K-means++ | Mean-shift | GMM | The method presented in this paper/ Terrace Compression Method + Window Function II + GMM |
|---|---|---|---|---|---|
| 1 | 0.3835 | **<u>0.3826</u>** | 0.4965 | **0.5468** | 0.4605 |
| 2 | **0.6793** | 0.6787 | 0.5964 | 0.6608 | **<u>0.5105</u>** |
| 3 | 0.7229 | 0.7221 | **<u>0.545</u>** | 0.8048 | 0.7128 |
| 4 | 0.7099 | 0.7092 | **<u>0.5053</u>** | 0.8674 | 0.7295 |
| 5 | 0.3875 | **<u>0.3868</u>** | 0.5001 | 0.5708 | 0.4293 |
| 6 | 0.6591 | 0.659 | 0.5824 | **0.684** | **<u>0.5129</u>** |
| 7 | 0.5781 | **_0.577_** | **0.8211** | 0.7299 | 0.6994 |
| 8 | 0.3838 | **_0.3824_** | **0.6456** | 0.6292 | 0.3837 |
| Overall Average | 0.563 | 0.5622 | 0.5866 | **0.6867** | _0.5548_ |

Note: Bold numbers indicate the maximum Dice value, while bold and underlined numbers indicate the minimum Dice value.

## 5. Conclusion

Breast cancer poses a threat to women's lives, and achieving routine detection methods is crucial for breast tumor prevention and improving patients' survival chances. In this background, detection based on multispectral transmission images offers clear advantages such as harmlessness, simplicity, and low cost, making it a promising tool for routine breast cancer screening. However, the signal-to-noise ratio (SNR) of breast transmission images is low, and frame accumulation techniques are used to enhance image quality, albeit leading to data redundancy. To address this issue, past research proposed the terraced compression method, achieving nonlinear compression of the grayscale of the image, yet requiring significant time and cost for manually tuning the dual thresholds. Therefore, this paper proposes a terraced compression method with automatic threshold selection based on grayscale distribution, combined with the Gaussian Mixture Model (GMM) clustering algorithm for detecting heterogeneous bodies. Initially, six-wavelength transmission spectral images of 16 heterogeneous bodies of different species with varying depths and thicknesses were collected. After enhancing the SNR of the images using frame accumulation and filtering techniques, the terraced compression method was employed to enhance image contrast and reduce data redundancy. Subsequently, grayscale window transformation was used to retain required information, and the images of each wavelength were composed into multi-dimensional image data for cluster analysis. Experimental results demonstrate the effective detection and classification of different substances in the mimic using the proposed method. This method effectively reduces the time and cost of debugging, presents a novel approach for clustering segmentation of fuzzy images, and lays the groundwork for the application of multispectral transmission imaging in early breast cancer screening.

## Acknowledgment

Thanks to the Key Laboratory of Biomedical Detecting Techniques of Tianjin University for the equipment and support provided.

## Funding

This research did not receive any specific grant from funding agencies in the public, commercial, or not-for-profit sectors.

## Conflict of Interest

The authors declare that they have no known competing financial interests or personal relationships that could have appeared to influence the work reported in this paper.